\documentclass[11pt]{article}

\usepackage{indentfirst}
\usepackage[utf8]{inputenc}
\usepackage{bm}
\usepackage{amsmath, amssymb, amsthm}
\usepackage{graphicx}
\usepackage{natbib}
\usepackage{geometry}
\usepackage{setspace}
\usepackage{hyperref}
\usepackage{booktabs}
\usepackage{algorithm}
\usepackage{algorithmic}
\usepackage{caption}
\usepackage{float}     
\usepackage{geometry}  
\usepackage{natbib}
\hypersetup{colorlinks=true, linkcolor=blue, citecolor=blue, urlcolor=blue}

\title{Functional Multidimensional Scaling}
\author{Liting Li}
\date{April 2025}

\begin{document}

\maketitle
\newpage
\begin{abstract}
This article introduces a functional method for lower-dimensional smooth representations in terms of time-varying dissimilarities. The method incorporates dissimilarity representation in multidimensional scaling and smoothness approach of functional data analysis by using cubic B-spline basis functions. The model is designed to arrive at optimal representations with an iterative procedure such that dissimilarities evaluated by estimated representations are almost the same as original dissimilarities of objects in a low dimension which is easier for people to recognize. To solve expensive computation in optimization, we propose a computationally efficient method by taking gradient steps with respect to individual sub-functions of target functions using a Stochastic Gradient Descent algorithm.

\end{abstract}

\newpage

\section{Introduction}
\noindent The development of analyzing positions of objects has been prevalent in cluster analysis and dimensionality reduction with complicated data. Multidimensional scaling (MDS) is one of the critical methods that provides a visual representation of proximities (similarities, dissimilarities, or distances) among a set of objects. For simplicity, in this work we only focus on dissimilarities. The main goal of MDS is to represent objects in a low-dimensional space without changing their original dissimilarities. MDS starts with dissimilarities instead of a feature representation. It rearranges objects and induces a space from dissimilarities. Constructing a low-dimensional space for objects is a major difference between MDS and many other dimensional reduction methods (for example, Principal Components Analysis). In addition to representing relationships between objects, MDS is also used to create a low-dimensional space (usually one, two, or three dimensions) for objects to avoid intractable computation and complex visual perception.\\
\indent However, dissimilarities are not always static in real situations. For example, the dissimilarities of closing prices on the S\&P 500 stocks can be different over time due to the change in closing prices. To date, there has been some research investigating the feasibility of temporal MDS methods. \cite{ambrosi1987dynamischer} designed a dynamic MDS method regarding dissimilarities measured in consecutive time periods. As one of the applications of time-varying MDS in the real world, \cite{machado2011analysis} conducted MDS analysis in 15 stock markets over the world with their time-varying correlations. Most recent research also includes the work studied by \cite{Jackle2016TMDS}, which is a temporal MDS visualization technique that creates one-dimensional MDS plots for multivariate time series data in network security. While temporal MDS is a growing field, far too little attention has been paid to smoothness of MDS representations if dissimilarities of objects are considered to change over time. Considering smooth MDS representations are worth to be analyzed, we take advantage of the techniques in Functional Data Analysis (FDA). FDA refers to an important field of statistics that studies samples of functions that are usually determined by smoothing raw data. It enables us to analyze data over a continuum, such as time, age, height, weight, wavelength, etc. Thus, objects in FDA are functions instead of numbers. Recent core studies on FDA are done by \cite{ramsay2002applied} and \cite{ramsay2005functional}.\\
\indent In order to study the smoothness of time-varying dissimilarities between objects, we produce a functional multidimensional scaling model with the B-spline basis to construct optimal low-dimensional configuration for objects. This paper is mainly composed of four sections. Firstly, we briefly review FDA (Section 2). Then we introduce the definition of dissimilarity and a detailed mathematical background in Classical MDS (CMDS), and we make a variant of CMDS with the closing prices of the S\&P 500 stocks from the first two weeks of the year 2018 to better understand the application of CMDS (Section 3). To extend CMDS by considering smoothness, we propose the functional multidimensional scaling (FMDS) method for time-varying data. Under the modified Adam Stochastic Gradient Descent that reduces the computation burden, the optimal smooth and low-dimensional representations can be estimated (Section 4). We illustrate the difference between FMDS and the existing MDS methods, and study its limitations and future work (Section 5).
\section{Background on Functional Data Analysis}
\noindent Functional data consists of independent and identically distributed curves (functions) over a continuous interval. According to the description made by \cite{ramsay2002applied}, we observe a set of $n$ objects consisting of the raw data $(y_{ij}, t_j)$ for $i = 1,...,n$ and $j = 1,...,m$, where $y_{ij}$ represents the value of the $i^{\text{th}}$ object at the time point $t_j$, assuming that the continuous interval mentioned previously is a time grid. In practice, the observed data is usually contaminated by random noise, resulting from random fluctuations around a smooth trajectory. \cite{ramsay2002applied} also pointed out that smoothness is the significant assumption in functional data analysis (FDA) so that the consecutive points $(y_{ij}, t_j)$ and $(y_{i,j+1},t_{j+1})$ are related to some extent and do not differ much from each other. There are two main types of functional data, dense functional data and sparse functional data. In this work, we focus on the dense functional data and start from the spine functions. Dense functional data is recorded on the same dense time grid of ordered times $\{t_1,...,t_m\}$ and the time grid is equally spaced, that is, $t_j - t_{j-1} = t_{j+1} - t_j$. When $m$ goes to infinity, $t_{j+1}-t_j$ tends to zero.

\subsection{Overview of Spline Functions}
Suppose that there is a grid (\cite{atkinson1989introduction})
\begin{center}
$a = t_0 < t_1 <...< t_{L+1} = b$.
\end{center}
$f(t)$ is a spline function of order $s \geq 1$ if it satisfies the following two properties:\\
(Property 1) $f(t)$ is a polynomial of degree $<s$ on each subinterval $[t_{i-1},t_i]$, for $i=1,...,L+1$.
(Property 2) The $d^{\text{th}}$ derivative $f^{\text{d}}(t)$ is continuous on $[a,b]$, for $0 \leq d \leq s-2$.

A spline divides into $L+1$ subintervals separated by values $t_l$, $l=1,...,L$ which are called knots. The order of a polynomial is the number of constants. Order 2, 3, and 4 are the most common situations in polynomials. The relation between the number of parameters $q$ in a spline function (the number of degrees of freedom), the order of the polynomials $s$ and the number of knots $L$ that was described by \cite{ramsay2005functional} indicates
\begin{center}
$q = s + L$.
\end{center}
For instance, the total number of parameters in the linear spline function is $q = 2 + 3 = 5$. A spline function $f(t)$ can be written as
\begin{center}
\[
f(t) = \sum_{k=1}^{q} c_k \phi_k(t),
\]
\end{center}
where $\phi_k(t)$'s are basis functions chosen from different basis systems, and $c_k$'s are coefficients which can be obtained by the least squares method in Section 2.2. B-spline basis system developed by \cite{deboor2001practical} is one of the most powerful basis systems.

Suppose that there exists a knot sequence $\{t_l\}$, the B-spline of order 1 for this knot sequence is defined as
\begin{center}
\[
B_{l1}(t) = \begin{cases}
1, & \text{if } t_l \leq t < t_{l+1} \\
0, & \text{otherwise}
\end{cases}
\]
\end{center}
with a constraint
\begin{center}
\[
\sum_{l}^{} B_{l1}(t) = 1,
\]
\end{center}
for all $t$.
When $t_l = t_{l+1}$, we have $B_{l1} = 0$. Based on the first-order B-splines, we obtain $s^{\text{th}}$-order $(s > 1)$ B-splines by recurrence
\begin{center}
$B_{ls} = w_{ls}B_{l,s-1} + (1 - w_{l+1,s})B_{l+1,s-1}$,
\end{center}
where 
\[
w_{ls}(t) = \begin{cases}
\frac{t-t_l}{t_{s+l-1}-t_l}, & \text{if } t_l \neq t_{l+s-1} \\
0, & \text{otherwise}
\end{cases}
\]
The cubic B-spline function (with order $s=4$) is commonly used as the basis functions $\phi_k(t)$.

\subsection{Smoothing Functional Data}
Suppose that there is an observation of the sample functions containing the raw data $\{(y_j, t_j)\}$, for $j = 1,...,m$, where $y_j$ represents the value observed at the time point $t_j$ that belongs to a compact interval $\tau$. In practice, the measurement process gives rise to random noise, thus, \cite{ramsay2005functional} expressed each of the actual $y_j$ data as 
\begin{center}
$y_j = x(t_j) + \varepsilon_j$,
\end{center}
where $x(t)$ is the smooth function in the stochastic process with the mean function $\mu(t) = \mathbb{E}(x(t))$ and covariance function $c(s,t) = Cov[x(s), x(t)]$, if $\mathbb{E}(\int_{t \in \tau}^{} x^2(t) \ dt) < \infty$, and $\varepsilon_j$ is the independent and identically distributed $(i,i,d)$ random noise with zero mean and constant variance $\sigma^2$.

As mentioned in Section 2.1, a spline function can be written as linear combinations of basis functions. As a useful tool to smooth data, the cubic B-spline function can be applied to $x(t)$. Accordingly, $x(t)$ can be modeled by
\begin{center}
\[
x(t) = \sum_{k=1}^{q} c_k \phi_k(t) = \mathbf{c}^\top \boldsymbol{\phi}(t),
\]
\end{center}
where $\phi_k(t)$'s are the cubic B-spline basis functions and $\mathbf{c}$ is a vector of length $q$ consisting the coefficients $c_k$'s, and $q$ depends on the order of the polynomials and the number of knots $L$, that is, $q = 4+L$. In order to estimate the smooth function $x(t)$, we need to determine the coefficient vector of $\mathbf {c}$ by minimizing the least squares
\[
SMSSE(\mathbf{y}|\mathbf{c}) = \sum_{j=1}^{m} [y_j - \sum_{k=1}^{q} c_k \phi_k(t_j)]^2.
\]
This can be done as follows.

Let $\mathbf{\Phi}$ be the $m \times q$ matrix containing the values of $\phi_k(t_j)$'s, for $j=1,...,m$, then
\begin{align*}
SMSSE(\mathbf{y}|\mathbf{c}) &= (\mathbf{y} - \mathbf{\Phi} \mathbf{c})^\top (\mathbf{y} - \mathbf{\Phi} \mathbf{c})\\
 &= \mathbf{y}^\top \mathbf{y} - \mathbf{c}^\top \mathbf{\Phi}^\top \mathbf{y} - \mathbf{y}^\top \mathbf{\Phi} \mathbf{c} + \mathbf{c}^\top \mathbf{\Phi}^\top \mathbf{\Phi} \mathbf{c},
\end{align*}
where $\mathbf{y} = [y_1,...,y_m]^\top$. By taking the partial derivative of $SMSSE(\mathbf{y}|\mathbf{c})$ with respect to $\mathbf{c}$, we have
\begin{align*}
\frac{\partial}{\partial \mathbf{c}}(\mathbf{y}^\top \mathbf{y} - \mathbf{c}^\top \mathbf{\Phi}^\top \mathbf{y} - \mathbf{y}^\top \mathbf{\Phi} \mathbf{c} + \mathbf{c}^\top \mathbf{\Phi}^\top \mathbf{\Phi} \mathbf{c}) &= 0 - \mathbf{\Phi}^\top \mathbf{y} - \mathbf{\Phi}^\top \mathbf{y} + 2 \mathbf{\Phi}^\top \mathbf{\Phi} \mathbf{c}\\
 &= 2 \mathbf{\Phi}^\top \mathbf{\Phi} \mathbf{c} - 2 \mathbf{\Phi}^\top \mathbf{y}.
\end{align*}
Let the above partial derivative be zero, we can estimate $\hat{\mathbf{c}}$ by
\begin{center}
$\hat{\mathbf{c}} = (\mathbf{\Phi}^\top \mathbf{\Phi})^{-1} \mathbf{\Phi}^\top \mathbf{y}$,
\end{center}
then the vector of fitted values is obtained by
\begin{center}
$\hat{\mathbf{y}} = \hat{x}(\mathbf{t}) = \mathbf{\Phi}\hat{\mathbf{c}} = \mathbf{\Phi}(\mathbf{\Phi}^\top \mathbf{\Phi})^{-1} \mathbf{\Phi}^\top \mathbf{y}$.
\end{center}

Since $\mathbf{\Phi}$ is an $m \times q$ matrix, it depends on the number of knots $L$ on the splines since $q=4+L$. On the other hand, there are various methods to decide $L$, but there is no gold standard methods for choosing the number of knots $L$.
\section{Background on Multidimensional Scaling}
\subsection{Dissimilarity and Dissimilarity Matrix}
\noindent Given that a set of $n$ objects and dissimilarities are metric distances (quantitative values) from \cite{izenman2008multidimensional}, let $d_{ij}$ represents the dissimilarity between the $i^{th}$ and the $j^{th}$ object, satisfying (a) $d_{ij} \geq 0$ for all $i,j$; (b) $d_{ii} = 0$ for all $i,j$; (c) $d_{ij} = d_{ji}$ for all $i,j$; (d) $d_{ij} \leq d_{is} + d_{sj}$ for all $i,j,s$. Dissimilarity can be measured as any kinds of distance metric, for example, we use the Euclidean distance to measure dissimilarity for $n$ objects $\mathbf{y}_1,...,\mathbf{y}_n \in \mathbb{R}^r$, then the dissimilarity between $\mathbf{y}_i = [y_{i1},y_{i2},...,y_{ir}]^\top$ and $\mathbf{y}_j = [y_{j1},y_{j2},...,y_{jr}]^\top$ for any $i,j=1,...n$ and $i \neq j$ is defined as \\
\[
d_{ij} = \|\mathbf{y}_i - \mathbf{y}_j\| = \{\sum_{k=1}^{r} (y_{ik}-y_{jk})^2\}^{1/2},
\]
and the squared Euclidean interpoint distance is evaluated by
\begin{center}
$d_{ij}^2 = \|\mathbf{y}_i - \mathbf{y}_j\|^2 = (\mathbf{y}_i - \mathbf{y}_j)^\top (\mathbf{y}_i - \mathbf{y}_j)$
\end{center}

The dissimilarities $\{d_{ij}\}$ construct an $(n \times n)$ dissimilarity matrix $\boldsymbol \Delta = (d_{ij})$. $\boldsymbol \Delta$ can be displayed as a lower-triangular or upper-triangular matrix because the diagonal entries are all zeros and it is a symmetric matrix.

Usually, the original data points $\mathbf{y}_1,...,\mathbf{y}_n \in \mathbb{R}^r$ are not given, instead, the dissimilarity matrix $\boldsymbol \Delta = (d_{ij})$ of these $n$ objects is provided. The aim of MDS is to find the optimal $p$-dimensional representations, $\mathbf{x}_1,...,\mathbf{x}_n \in \mathbb{R}^p \ (p<r)$, such that the dissimilarity $\delta_{ij}$ between $\mathbf{x}_i$ and $\mathbf{x}_j$ satisfies
\begin{center}
$\delta_{ij} \approx d_{ij}$.
\end{center}
It means that the dissimilarity of the optimal $p$-dimensional representation, $\mathbf{x}_1,...,\mathbf{x}_n \in \mathbb{R}^p$ preserves the dissimilarities of the original data points $\mathbf{y}_1,...,\mathbf{y}_n \in \mathbb{R}^r$.

\subsection{Classical Multidimensional Scaling}
Classical multidimensional scaling (CMDS) is one of the main methods in multidimensional scaling and was first introduced by \cite{young1938discussion}. In CMDS, there are three conditions required: (1) let $\mathbf{x}_i = (x_{i1},x_{i2},...,x_{ip})^\top$, we assume that
\[
\sum_{i=1}^{n}x_{ik} = 0, \quad \text{for all} \ k=1,2,...,p; \tag{3.1}
\]
(2) the dissimilarities are measured by the Euclidean distances; (3) the parametric function $f(d_{ij}) = d_{ij}$. Thus, the purpose of CMDS is to find a configuration of points $\mathbf{x}_1,...,\mathbf{x}_n \in \mathbb{R}^p \ (p<r)$ such that $\|\mathbf{x}_i - \mathbf{x}_j\| = d_{ij}$. According to the description in \cite{ramsay2005functional}, the solution of CMDS problem is not unique. It can be proved by considering an orthogonal matrix $\boldsymbol{\Gamma}$ and an arbitrary vector $\mathbf{c}$ that form a transformation of $\mathbf{x}_i$ and $\mathbf{x}_j$: $\mathbf{x}_i \rightarrow \boldsymbol{\Gamma} \mathbf{x}_i + \mathbf{c}$ and $\mathbf{x}_j \rightarrow \boldsymbol{\Gamma} \mathbf{x}_j + \mathbf{c}$. Thus, we have 
\begin{align*}
\|(\boldsymbol{\Gamma} \mathbf{x}_i + \mathbf{c}) - (\boldsymbol{\Gamma} \mathbf{x}_j + \mathbf{c})\|^2 &= \|\boldsymbol{\Gamma}(\mathbf{x}_i - \mathbf{x}_j)\|^2 \\
  &= (\mathbf{x}_i - \mathbf{x}_j)^\top \boldsymbol{\Gamma}^\top \boldsymbol{\Gamma} (\mathbf{x}_i - \mathbf{x}_j) \\
  &= \|\mathbf{x}_i - \mathbf{x}_j \|^2,
\end{align*}
which indicates that a common orthogonal transformation causes different solution of CMDS problem. One way of determining the solution is to look at the eigenvalues of $\mathbf{B}$ in CMDS algorithm that was also described in \cite{ramsay2005functional}.

In CMDS algorithm, $b_{ij}=\mathbf{x}_i^\top \mathbf{x}_j \ (i,j=1,...,n)$ are used as the elements to form an $n \times n$ symmetric matrix $\mathbf{B}$. 
Define an $(n \times n)$ matrix $\mathbf{A} = (a_{ij})$ with the elements,
$a_{ij} = -\frac{1}{2}d_{ij}^2$, $a_{i.} = n^{-1} \sum\limits_{j} a_{ij}$, $a_{.j} = n^{-1} \sum\limits_{i} a_{ij}$, $a_{..} = n^{-2} \sum\limits_{i} \sum\limits_{j} a_{ij}$. The matrix $\mathbf{B}$ is given by
\begin{center}
$\mathbf{B} = \mathbf{HAH}$,
\end{center}
where $\mathbf{H} = \mathbf{I}_n - n^{-1} \mathbf{J}_n$ and $\mathbf{J}_n = \mathbf{1}_n \mathbf{1}_n^\top$ is an $n \times n$ matrix of ones.

If $\mathbf{B}$ is positive semi-definite with rank $r(\mathbf{B})=p<n$, then the largest $p$ eigenvalues $\lambda_1,...,\lambda_p$ are positive and the remaining $n-p$ eigenvalues are zeros. The largest $p$ eigenvalues form a matrix
\[
\mathbf{\Lambda_1} = 
\begin{bmatrix}
\lambda_1 & 0 & 0 & \cdots & 0 \\
0 & \lambda_2 & 0 & \cdots & 0 \\
0 & 0 & \lambda_3 & \cdots & 0 \\
\vdots & \vdots & \vdots & \ddots & \vdots \\
0 & 0 & 0 & \cdots & \lambda_p
\end{bmatrix}.
\]
Assume that $\mathbf{v}_1,...,\mathbf{v}_p$ are the corresponding eigenvectors and $\mathbf{V}_1 = (\mathbf{v}_1,...,\mathbf{v}_p)$, then we can rewrite $\mathbf{B}$ as
\begin{center}
$\mathbf{B} = \mathbf{V}_1 \mathbf{\Lambda}_1 \mathbf{V}_1^\top = (\mathbf{V}_1 \mathbf{\Lambda}_1^{1/2})(\mathbf{\Lambda}_1^{1/2} \mathbf{V}_1^\top)$.
\end{center}
Since $b_{ij}=\mathbf{x}_i^\top \mathbf{x}_j$, $\mathbf{B}$ can also be expressed as $\mathbf{B} = \mathbf{X} \mathbf{X}^\top$, where $\mathbf{X} = [\mathbf{x}_1,...,\mathbf{x}_n]^\top$. Hence, the matrix of optimal solutions is 
\begin{center}
$\mathbf{\hat X} = \mathbf{V}_1 \mathbf{\Lambda}_1^{1/2} = (\sqrt{\lambda_1} \mathbf{v}_1,...,\sqrt{\lambda_p} \mathbf{v}_p) = (\mathbf{\hat x}_1,...,\mathbf{\hat x}_n)^\top$,
\end{center}
where $\mathbf{x}_i$ is a $(p \times 1)$-vector for all $i=1,...,n$. As a result, $\mathbf{\hat x}_1,...,\mathbf{\hat x}_n$ produce a $p$-dimensional space and are the optimal solutions to the representation in CMDS.

\subsection{Application of Variant Classical Multidimensional Scaling}

As mentioned in the section 3.1, dissimilarities do not have to be the Euclidean distances. In this use case, we apply the algorithm of CMDS to the daily closing prices of the S\&P 500 stocks, but compute the dissimilarities with the correlations instead of the Euclidean distances. In a given week, we observe the daily closing prices of stocks $i$ and $j$, which are $y_{ik}$ and $y_{jk}$, for $i,j=1,2,...,500$, $i \neq j$, and $k=1,...,r$, where $r$ is the number of trading days in the week. Thus, their correlation $R_{ij}$ in the week is given by
\[
R_{ij} = \frac{\sum_{k=1}^{r}(y_{ik}-\bar{y}_i)(y_{jk}-\bar{y}_j)}{\sqrt{\sum_{k=1}^{r}(y_{ik}-\bar{y}_i)^2} \sqrt{\sum_{k=1}^{r}(y_{jk}-\bar{y}_j)^2}} ,
\]
where $\bar{y}_i = r^{-1} \sum_{k=1}^{r} y_{ik}$ and $\bar{y}_j = r^{-1} \sum_{k=1}^{r} y_{jk}$.

Define the dissimilarities of the closing prices as 
\begin{center}
$d_{ij}= \frac{1-R_{ij}}{2}$ .
\end{center}
Thus, the dissimilarity matrix $\mathbf{D}$ can be formed with $\{d_{ij}\}$'s
\[
\mathbf{D} = 
\begin{bmatrix}
0 & d_{12} & \cdots & d_{1,500} \\
d_{21} & 0 & \cdots & d_{2,500} \\
\vdots & \vdots & \ddots & \vdots \\
d_{500,1} & d_{500,2} & \cdots & 0
\end{bmatrix},
\]
where
$ 
\begin{cases}
d_{ij} = d_{ji}, & \text{for } i \neq j, \\
d_{ii} = 0, & \text{for all } i.
\end{cases}
$

Consider the first trading week of the year 2018 and implement CMDS method with the above dissimilarity matrix $\mathbf{D}$, we obtain the 2-dimensional $(p=2)$ and 3-dimensional $(p=3)$ plots of the stock closing prices in MDS reconstruction. Each plot has 500 points as one point $\mathbf{\hat x}_i$ represents one stock. 

\begin{figure}[H]
    \centering
    \includegraphics[width=0.8\linewidth]{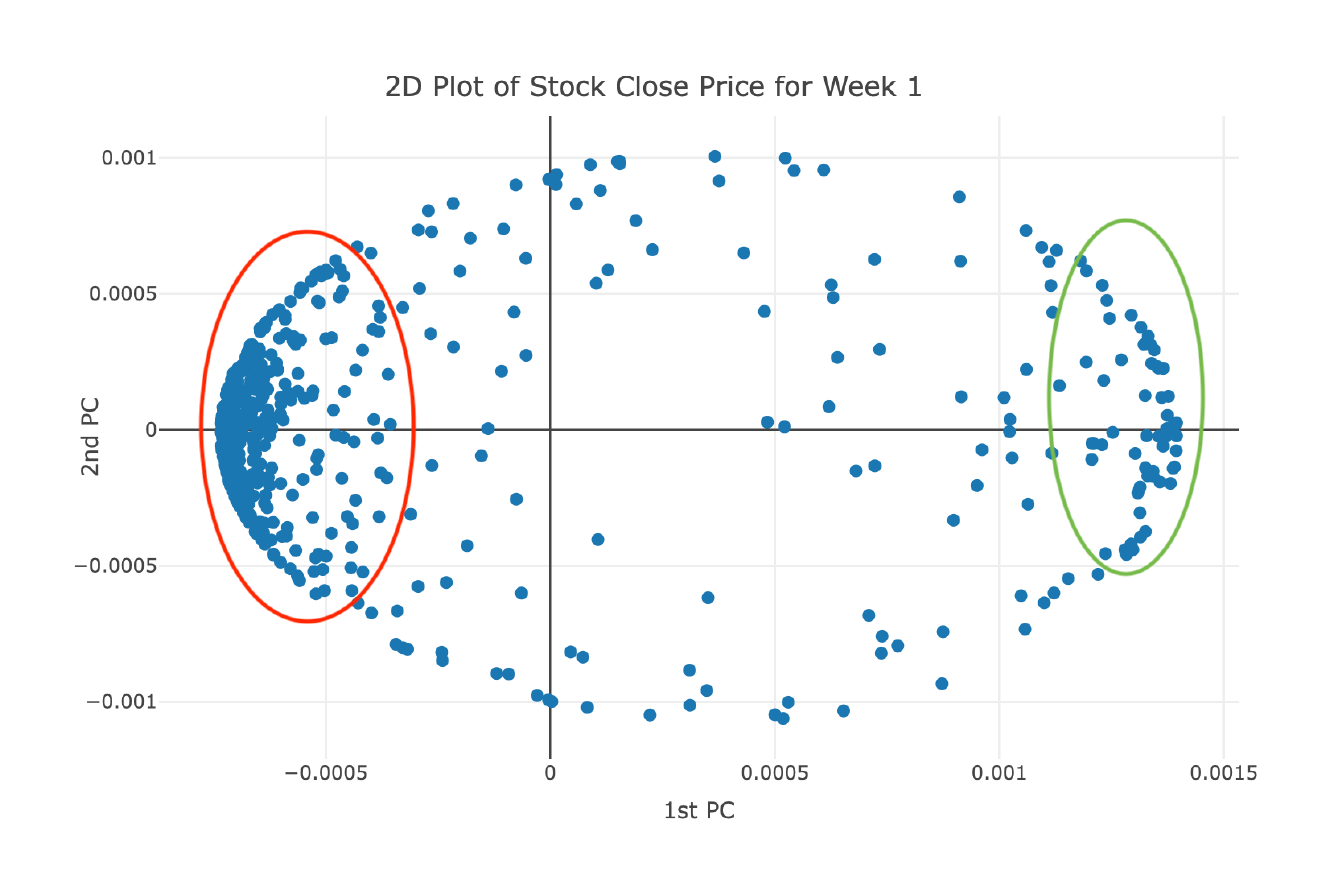}
    \captionsetup{font=small}
    \caption{2D MDS Plot of the S\&P Stocks Closing Prices in the $1^{st}$ Week of Year 2018}
    \label{fig:enter-label}
\end{figure}

Figure 1 displays the plot of the 2D CMDS solution. Although some points are sparsity, there are two obvious clusters of the stocks which are circled by red and green. The red cluster has much more stocks than the green cluster. The points which are closer together indicate the corresponding stocks having smaller dissimilarities (larger correlation). In contrast, the points which are far apart from each other indicate the corresponding stocks having large dissimilarities (small correlation). But there are some stocks in the middle part not having obvious cluster which means that they have large dissimilarities from the others. We expect the same situation in the 3D MDS solution for this week.

\begin{figure}[H]
    \centering
    \includegraphics[width=0.8\linewidth]{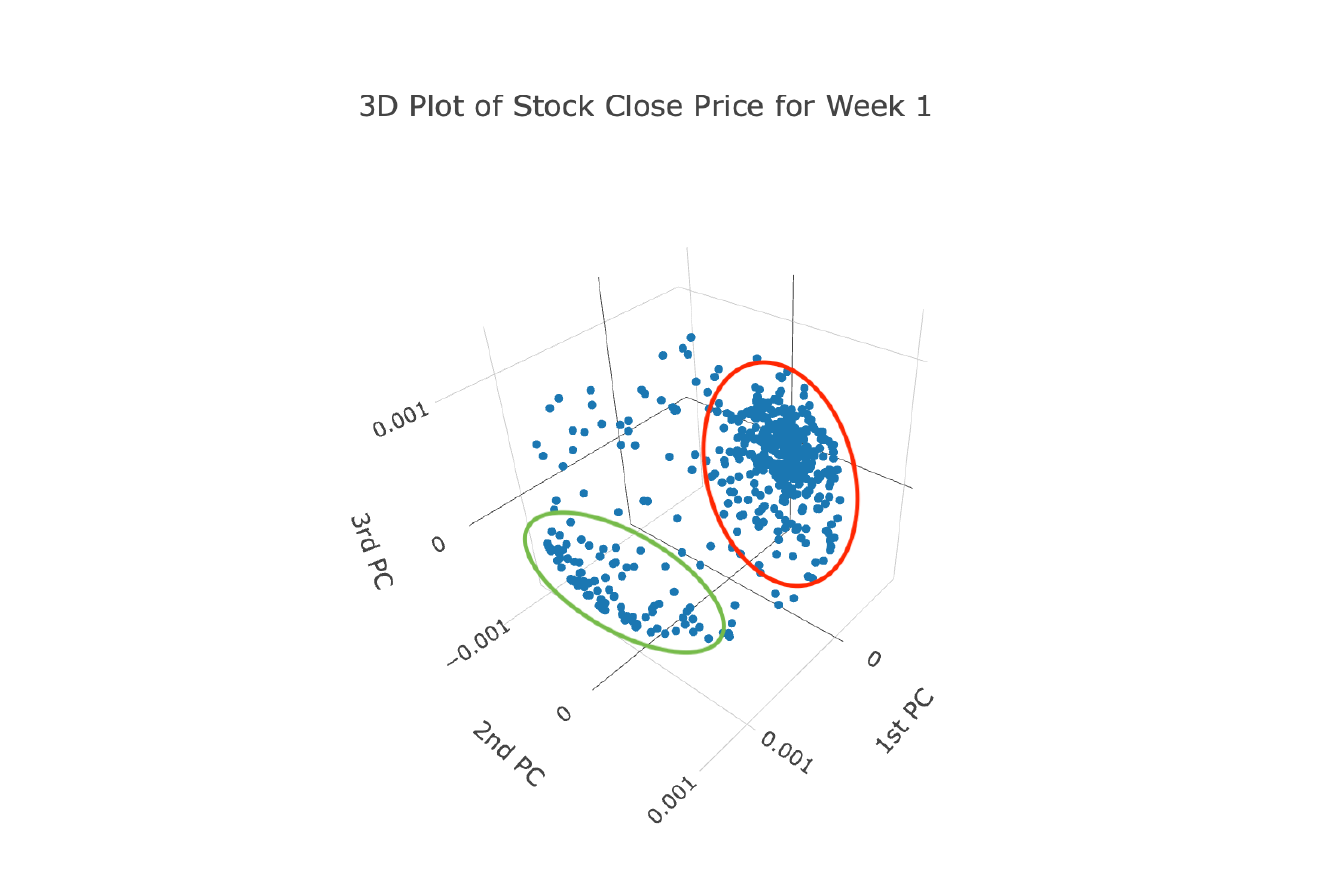}
    \captionsetup{font=small}
    \caption{3D MDS Plot of the S\&P Stocks Closing Prices in the $1^{st}$ Week of Year 2018}
    \label{fig:enter-label}
\end{figure}

Figure 2 is the corresponding 3D CMDS reconstruction and shows the same situation as Figure 1. There are also two obvious clusters (red and green), where the red cluster has much more stocks than the green cluster. But the rest of the stocks do not have obvious clusters which means they have large dissimilarities from the others. 

In these two plots, we can see that CMDS provides a good visualization of the closing prices of the S\&P 500 stocks in a fixed week. Furthermore, we want to see how the situation changes from the first trading week to the second trading week of the year 2018. Thus, we create the 2-dimensional $(p=2)$ and 3-dimensional $(p=3)$ plots for the second trading week as shown in Figure 3 and 4, respectively.

\begin{figure}[H]
    \centering
    \includegraphics[width=0.8\linewidth]{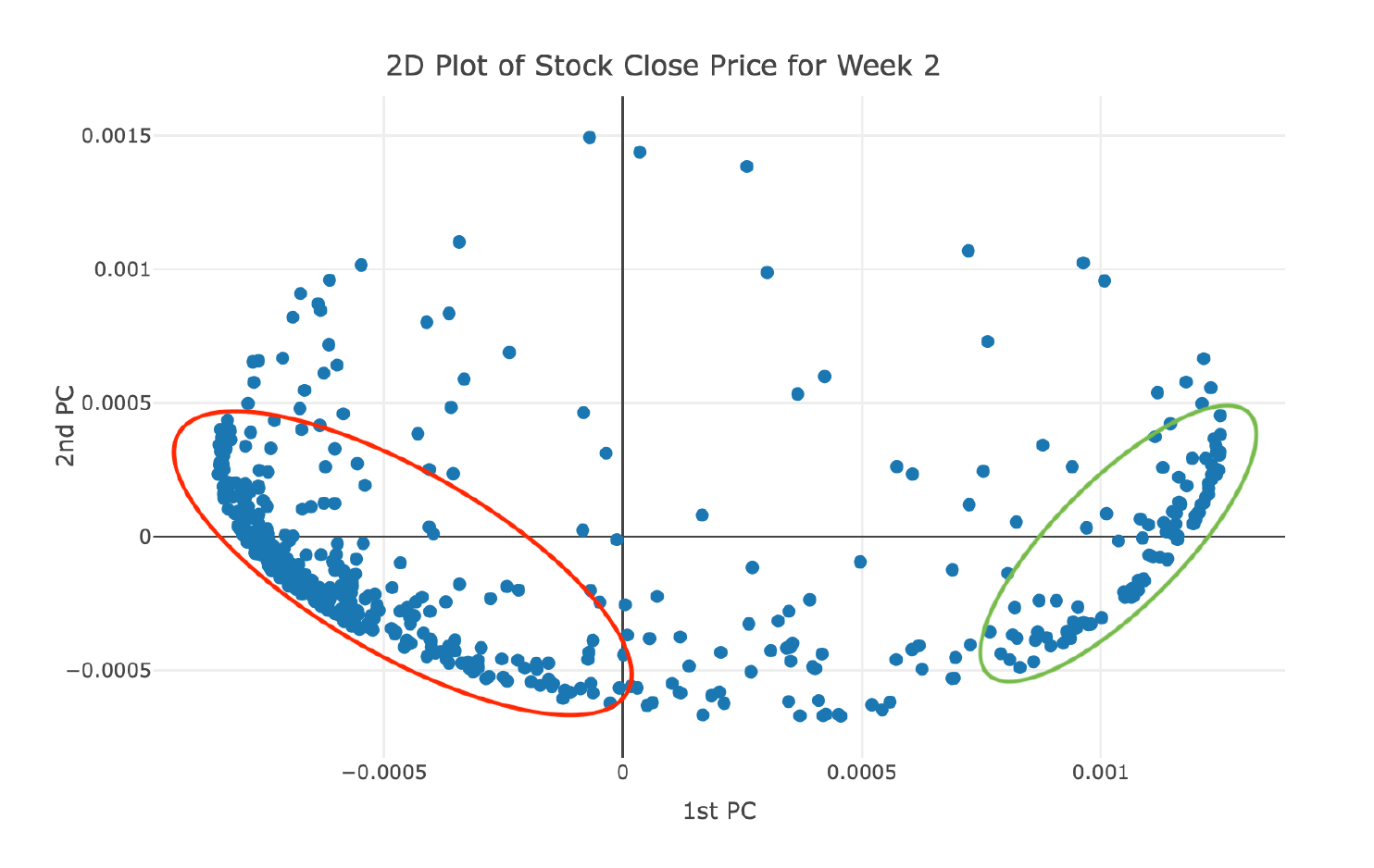}
    \captionsetup{font=small}
    \caption{2D MDS Plot of the S\&P Stocks Closing Prices in the $2^{nd}$ Week of Year 2018}
    \label{fig:enter-label}
\end{figure}

\begin{figure} [H]
    \centering
    \includegraphics[width=0.8\linewidth]{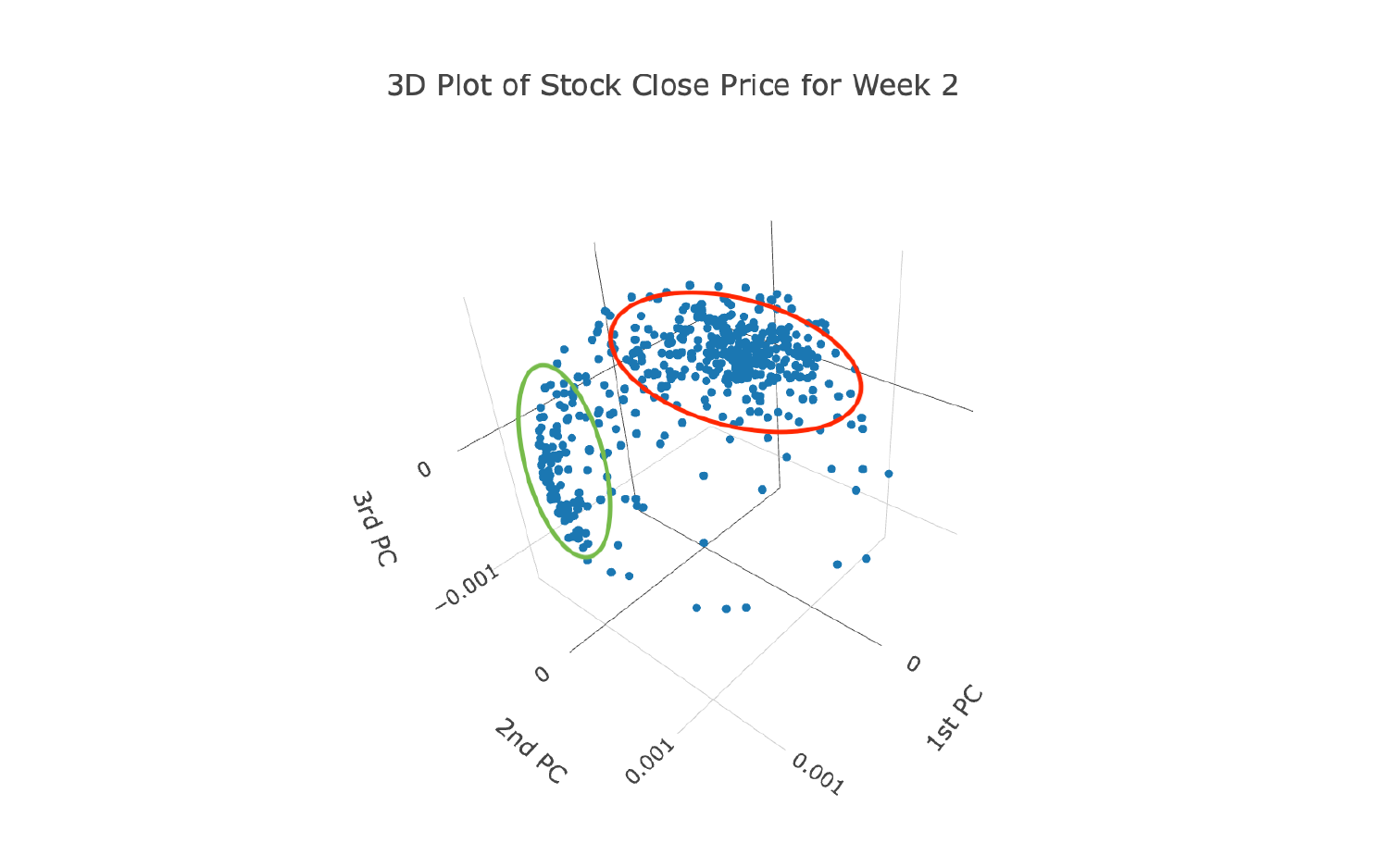}
    \captionsetup{font=small}
    \caption{3D MDS Plot of the S\&P Stocks Closing Prices in the $2^{nd}$ Week of Year 2018}
    \label{fig:enter-label}
\end{figure}

Although the pattern of the points seems different from the first week to the second week, the clusters actually do not change much from week to week. The two different vertical axes make the different visual effects somehow. In general, the plots are still similar. The stocks are mostly separated into two obvious clusters (circled by red and green) in which one cluster has much more stocks than the other cluster.

However, the above MDS solutions are only evaluated by the fixed and static time (week), and we do not obtain the MDS configuration changing over time. Also, $\mathbf{\hat x}_i$'s are constructed from the eigenvectors. There is no guarantee that the eigenvectors change smoothly from time $t_k$ to $t_{k+1}$, for each time point $t$ in a continuous interval $\tau$. In other words, the above MDS solutions are discrete, but not continuous during a time interval. In order to solve this issue and provide a smooth MDS representation over time, we propose a new method in the next section.
\section{Functional Multidimensional Scaling}
\noindent In order to obtain smoothly time-varying representation $\mathbf{x}_i(t)$'s, we come up with a functional multidimensional scaling method (FMDS) such that
\begin{center}
$\|\mathbf{x}_i(t) - \mathbf{x}_j(t)\| \approx d_{ij}(t)$, for $i,j=1,...,n$,
\end{center} 
where $d_{ij}(t)$ is the dissimilarities between the given objects $\mathbf{y}_i(t)$ and $\mathbf{y}_j(t)$. Since B-spline basis functions contribute smoothness to functional data, we let $\mathbf{x}_i(t) = \mathbf{C} \boldsymbol{\beta}(t)$, where $\boldsymbol{\beta}(t)$ is the vector of $q$ B-spline basis functions, and each $\mathbf{C}_i$ is a $(p \times q)$ coefficient matrix, for $i=1,...,n$. We can estimate $\mathbf{C}_1,...,\mathbf{C}_n$ by minimizing the target function
\begin{align*}
F(\mathbf{C}_1,...,\mathbf{C}_n) &= \sum_{i<j}^{n} \sum_{k=1}^{m} [d_{ij}^2(t_k) - \|\mathbf{C}_i \boldsymbol{\beta}(t_k) - \mathbf{C}_j \boldsymbol{\beta}(t_k)\|^2]^2 \\
 &= \sum_{i<j}^{n} \sum_{k=1}^{m} [d_{ij}^2(t_k) - (\mathbf{C}_i \boldsymbol{\beta}(t_k) - \mathbf{C}_j \boldsymbol{\beta}(t_k))^\top (\mathbf{C}_i \boldsymbol{\beta}(t_k) - \mathbf{C}_j \boldsymbol{\beta}(t_k))]^2, \tag{4.1}
\end{align*}
where $m$ is the length of time period with a continuous interval $\tau$.

The minimization can be achieved by letting the partial derivatives of $\mathbf{C}_h$ be zero for $h \neq j$, where $j=1,...,n$, that is, $\frac{\partial F}{\partial\mathbf{C}_h} = \mathbf{0}$ which is a $(p \times q)$ matrix. Thus, we have the partial derivatives as follows \\
\begin{align*}
\frac{\partial F}{\partial{\mathbf{C}_h}} &= \sum_{\substack{j=1 \\ j \neq h}}^{n} \sum_{k=1}^{m} \frac{\partial}{\partial \mathbf{C}_h}[d_{hj}^2(t_k) - (\mathbf{C}_h \boldsymbol{\beta}(t_k) - \mathbf{C}_j \boldsymbol{\beta}(t_k))^\top (\mathbf{C}_h \boldsymbol{\beta}(t_k) - \mathbf{C}_j \boldsymbol{\beta}(t_k))]^2 \\
  &= -4\sum_{\substack{j=1 \\ j \neq h}}^{n} \sum_{k=1}^{m} [d_{hj}^2(t_k) - (\mathbf{C}_h \boldsymbol{\beta}(t_k) - \mathbf{C}_j \boldsymbol{\beta}(t_k))^\top (\mathbf{C}_h \boldsymbol{\beta}(t_k) - \mathbf{C}_j \boldsymbol{\beta}(t_k))](\mathbf{C}_h \boldsymbol{\beta}(t_k) - \mathbf{C}_j \boldsymbol{\beta}(t_k)) \boldsymbol{\beta}(t_k)^\top
\end{align*}
Accordingly, we let
\begin{center}
\[
\sum_{\substack{j=1 \\ j \neq h}}^{n} \sum_{k=1}^{m} [d_{hj}^2(t_k) - (\mathbf{C}_h \boldsymbol{\beta}(t_k) - \mathbf{C}_j \boldsymbol{\beta}(t_k))^\top (\mathbf{C}_h \boldsymbol{\beta}(t_k) - \mathbf{C}_j \boldsymbol{\beta}(t_k))](\mathbf{C}_h \boldsymbol{\beta}(t_k) - \mathbf{C}_j \boldsymbol{\beta}(t_k)) \boldsymbol{\beta}(t_k)^\top = 0
\]
\end{center}

In the above equation, each $\mathbf{C}_h$ cannot be estimated in a closed form. To solve the parameters $\mathbf{C}_h$ out from the equation, we need to use an iterative procedure such as the Quasi-Newton method introduced by \cite{davidon1991variable}. Another difficulty in estimating the parameters $\mathbf{C}_h$ is that the computation can be time-consuming when the optimization problem is high-dimensional. For example, if $n=500$, $p=2$, and $q=10$, then the array $\mathbf{C} = (\mathbf{C}_1,...,\mathbf{C}_{500})$ has $500 \times 2 \times 10 = 10,000$ elements in total. Suppose that we take $1,000$ iterations, then there will be $10,000 \times 1,000 = 10,000,000$ computations to complete the algorithm. In order to reduce the computation burden, we came up with a method of stochastic gradient descent (SGD).

In practice, many target functions are composed of a sum of sub-functions, such as $F(\mathbf{C}_1,...,\mathbf{C}_n)$ in (4.1). SGD methods complete optimizations by taking gradient steps with respect to the individual sub-functions of target functions. Adam SGD proposed by \cite{kingma2015adam} is a type of SGD methods which only requires first-order gradients of individual sub-functions with little memory requirement so that it is efficient for the high-dimensional optimization. In our target function (4.1), the individual sub-functions are the functions corresponding to the $h^{th}$ and $j^{th}$ objects as follows
\begin{align*}
f(\mathbf{C}_h, \mathbf{C}_j) &= \sum_{k=1}^{m}[d_{hj}^2(t_k) - \|\mathbf{x}_h(t_k) - \mathbf{x}_j(t_k)\|^2]^2 \\
  &= \sum_{k=1}^{m} [d_{hj}^2(t_k) - (\mathbf{C}_h \boldsymbol{\beta}(t_k) - \mathbf{C}_j \boldsymbol{\beta}(t_k))^\top (\mathbf{C}_h \boldsymbol{\beta}(t_k) - \mathbf{C}_j \boldsymbol{\beta}(t_k))]^2,
\end{align*}
which results in the first-order stochastic gradient functions with respect to $\mathbf{C}_h$ and $\mathbf{C}_j$
\begin{align*}
\frac{\partial f}{\partial \mathbf{C}_h} &= -4\sum_{k=1}^{m} [d_{hj}^2(t_k) - (\mathbf{C}_h \boldsymbol{\beta}(t_k) - \mathbf{C}_j \boldsymbol{\beta}(t_k))^\top (\mathbf{C}_h \boldsymbol{\beta}(t_k) - \mathbf{C}_j \boldsymbol{\beta}(t_k))][\mathbf{C}_h \boldsymbol{\beta}(t_k) - \mathbf{C}_j \boldsymbol{\beta}(t_k)] \boldsymbol{\beta}(t_k)^\top, \\
\frac{\partial f}{\partial \mathbf{C}_j} &= 4\sum_{k=1}^{m} [d_{hj}^2(t_k) - (\mathbf{C}_h \boldsymbol{\beta}(t_k) - \mathbf{C}_j \boldsymbol{\beta}(t_k))^\top (\mathbf{C}_h \boldsymbol{\beta}(t_k) - \mathbf{C}_j \boldsymbol{\beta}(t_k))][\mathbf{C}_h \boldsymbol{\beta}(t_k) - \mathbf{C}_j \boldsymbol{\beta}(t_k)] \boldsymbol{\beta}(t_k)^\top.
\end{align*}
Assembled with Adam SGD, each iteration $i$ updates each pair of $\mathbf{C}_h$ and $\mathbf{C}_j$ with random sample $h$ selected from $\{1,...,n-1\}$ and $j=h+1,h+2,...,n$, respectively, that is, \\
\begin{align*}
\mathbf{C}_h^{(i+1)} &= \mathbf{C}_h^{(i)} - \alpha \frac{\mathbf{\hat m}_h^{(i+1)}}{\sqrt{\mathbf{\hat v}_h^{(i+1)}} + \mathbf{e}_h} ,\\
\mathbf{C}_j^{(i+1)} &= \mathbf{C}_j^{(i)} - \alpha \frac{\mathbf{\hat m}_j^{(i+1)}}{\sqrt{\mathbf{\hat v}_j^{(i+1)}} + \mathbf{e}_j} ,
\end{align*}
where $\alpha = 0.001$, $\mathbf{e}_h$ and $\mathbf{e}_j$ are both $p \times q$ constant matrices with all elements equal to $10^{-8}$, as for $\mathbf{\hat m}_h^{(i+1)}$, $\mathbf{\hat m}_j^{(i+1)}$, $\mathbf{\hat v}_h^{(i+1)}$, and $\mathbf{\hat v}_j^{(i+1)}$, they are given as follows.
\begin{align*}
\mathbf{\hat m}_h^{(i+1)} &= \frac{\mathbf{m}_h^{(i+1)}}{1-\gamma_1^{i+1}}, \\
\mathbf{\hat m}_j^{(i+1)} &= \frac{\mathbf{m}_j^{(i+1)}}{1-\gamma_1^{i+1}}, \\
\mathbf{\hat v}_h^{(i+1)} &= \frac{\mathbf{v}_h^{(i+1)}}{1-\gamma_2^{i+1}}, \\
\mathbf{\hat v}_j^{(i+1)} &= \frac{\mathbf{v}_j^{(i+1)}}{1-\gamma_2^{i+1}},
\end{align*}
where \\
\begin{align*}
\mathbf{m}_h^{(i+1)} &= \gamma_1 \mathbf{m}_h^{(i)} + (1-\gamma_1) \frac{\partial f}{\partial \mathbf{C}_h}, \\
\mathbf{m}_j^{(i+1)} &= \gamma_1 \mathbf{m}_j^{(i)} + (1-\gamma_1) \frac{\partial f}{\partial \mathbf{C}_j}, \\
\mathbf{v}_h^{(i+1)} &= \gamma_2 \mathbf{v}_h^{(i)} + (1-\gamma_2) \frac{\partial f}{\partial \mathbf{C}_h} \circ \frac{\partial f}{\partial \mathbf{C}_h},\\
\mathbf{v}_j^{(i+1)} &= \gamma_2 \mathbf{v}_j^{(i)} + (1-\gamma_2) \frac{\partial f}{\partial \mathbf{C}_j} \circ \frac{\partial f}{\partial \mathbf{C}_j}.
\end{align*}
In Adam SGD, $\alpha$ is the step length, $\gamma_1$ and $\gamma_2$ are the decay rate parameters. Typically, $\gamma_1 = 0.9$, $\gamma_2 = 0.999$. The initial $\mathbf{m}_h^{(0)} = \mathbf{0}$, $\mathbf{v}_h^{(0)} = \mathbf{0}$, $\mathbf{m}_j^{(0)} = \mathbf{0}$, $\mathbf{v}_j^{(0)} = \mathbf{0}$ are $p \times q$ matrices. It is important to note that the random sample of $h$ is without replacement from $\{1,...,n-1\}$ in each for loop. In order to speed up the convergence of the above algorithm, we initialize $\mathbf{C}_1,...,\mathbf{C}_n$ by using CMDS with the given dissimilarities $d_{ij}$'s.

The iteration stops when $\mathbf{C}_1,...,\mathbf{C}_n$ tend to converge. In general, the algorithm is described in Algorithm 1. The resulting parameters are denoted by $\mathbf{\hat C}_1,...,\mathbf{\hat C}_n$ such that the optimal $p$-dimensional FMDS configurations are derived by $\mathbf{\hat x}_1(t) = \mathbf{\hat C}_1 \boldsymbol{\beta}(t),...,\mathbf{\hat x}_n(t) = \mathbf{\hat C}_n \boldsymbol{\beta}(t)$.

\begin{algorithm}[H]
\caption{Modified Adam Stochastic Gradient Descent for FMDS}
\label{alg:modified_adam}
\begin{algorithmic}[1]
\REQUIRE Learning rate $\alpha = 0.001$, exponential decay rates $\gamma_1 = 0.9$, $\gamma_2 = 0.999$, \\
$\mathbf{e}_h$ and $\mathbf{e}_j$ are both $p \times q$ constant matrices with all elements equal to $10^{-8}$, \\
initial parameters $\mathbf{C}_h$ and $\mathbf{C}_j$ by using CMDS with the given dissimilarities $d_{ij}$'s, \\
$h$ is a random sample selected from $\{1,...,n-1\}$ and $j=h+1,h+2,...,n$.
\ENSURE Optimized each pair of parameters $\mathbf{C}_h$ and $\mathbf{C}_j$
\STATE Initialize $\mathbf{m}_h^{(0)} = \mathbf{0}$, $\mathbf{v}_h^{(0)} = \mathbf{0}$, $\mathbf{m}_j^{(0)} = \mathbf{0}$, $\mathbf{v}_j^{(0)} = \mathbf{0}$ are $p \times q$ matrices, iteration step $i = 0$
\WHILE{$\|\mathbf{C}_h^{(i+1)} - \mathbf{C}_h^{(i)}\| \geq \epsilon$ and $\|\mathbf{C}_j^{(i+1)} - \mathbf{C}_j^{(i)}\| \geq \epsilon$}
    \STATE Compute gradient $\frac{\partial f}{\partial \mathbf{C}_h^{(i)}}$ and $\frac{\partial f}{\partial \mathbf{C}_j^{(i)}}$
    \STATE Update $\mathbf{m}_h^{(i+1)} = \gamma_1 \mathbf{m}_h^{(i)} + (1-\gamma_1) \frac{\partial f}{\partial \mathbf{C}_h^{(i)}}$
    \STATE Update $\mathbf{m}_j^{(i+1)} = \gamma_1 \mathbf{m}_j^{(i)} + (1-\gamma_1) \frac{\partial f}{\partial \mathbf{C}_j^{(i)}}$
    \STATE Update $\mathbf{v}_h^{(i+1)} = \gamma_2 \mathbf{v}_h^{(i)} + (1-\gamma_2) \frac{\partial f}{\partial \mathbf{C}_h^{(i)}} \circ \frac{\partial f}{\partial \mathbf{C}_h^{(i)}}$
    \STATE Update $\mathbf{v}_j^{(i+1)} = \gamma_2 \mathbf{v}_j^{(i)} + (1-\gamma_2) \frac{\partial f}{\partial \mathbf{C}_j^{(i)}} \circ \frac{\partial f}{\partial \mathbf{C}_j^{(i)}}$
    \STATE Bias correction $\mathbf{\hat m}_h^{(i+1)} = \frac{\mathbf{m}_h^{(i+1)}}{1-\gamma_1^{i+1}}$
    \STATE Bias correction $\mathbf{\hat m}_j^{(i+1)} = \frac{\mathbf{m}_j^{(i+1)}}{1-\gamma_1^{i+1}}$
    \STATE Bias correction $\mathbf{\hat v}_h^{(i+1)} = \frac{\mathbf{v}_h^{(i+1)}}{1-\gamma_2^{i+1}}$
    \STATE Bias correction $\mathbf{\hat v}_j^{(i+1)} = \frac{\mathbf{v}_j^{(i+1)}}{1-\gamma_2^{i+1}}$
    \STATE Update parameters: $\mathbf{C}_h^{(i+1)} = \mathbf{C}_h^{(i)} - \alpha \frac{\mathbf{\hat m}_h^{(i+1)}}{\sqrt{\mathbf{\hat v}_h^{(i+1)}} + \mathbf{e}_h}$
    \STATE Update parameters: $\mathbf{C}_j^{(i+1)} = \mathbf{C}_j^{(i)} - \alpha \frac{\mathbf{\hat m}_j^{(i+1)}}{\sqrt{\mathbf{\hat v}_j^{(i+1)}} + \mathbf{e}_j}$
    \STATE Update $i = i + 1$
\ENDWHILE
\RETURN each pair of parameters $\mathbf{C}_h$ and $\mathbf{C}_j$
\end{algorithmic}
\end{algorithm}
\section{Discussion}
\noindent As a novel statistical framework, FMDS synthesizes the strengths of dense functional data analysis and the method of Classical MDS. By seamlessly integrating the smoothness approach in functional data analysis with the characteristic of preserving dissimilarity in Classical MDS, FMDS provides a way to identify optimal and low-dimensional space
\begin{center}
$\mathbf{\hat x}_1(t)=\mathbf{\hat C}_1 \boldsymbol{\beta}(t), \mathbf{\hat x}_2(t)=\mathbf{\hat C}_2 \boldsymbol{\beta}(t),..., \mathbf{\hat x}_n(t)=\mathbf{\hat C}_n \boldsymbol{\beta}(t)$
\end{center}
for $n$ objects where $n$ is fixed and addresses the key limitations previously observed in Classical MDS, particularly in scenarios involving dissimilarities measured in continuous time periods, complex data structures, smoothness of MDS solutions, and expensive computations.

For each $t_m$ in a dense time interval $\tau$, suppose that the dissimilarity of any pair of the $n$ objects is $d_{ij}(t_m)$ and the dissimilarity of any pair of optimal representations $\mathbf{\hat x}_i(t_m)=\mathbf{\hat C}_i \boldsymbol{\beta}(t_m)$ and $\mathbf{\hat x}_j(t_m)=\mathbf{\hat C}_j \boldsymbol{\beta}(t_m)$ ($i,j=1,2,...,n, \ i \neq j$) is evaluated by
\begin{center}
$\hat {d}_{ij}(t_m) = \|\mathbf{\hat x}_i(t_m) - \mathbf{\hat x}_j(t_m) \| = \|\mathbf{\hat C}_i \boldsymbol{\beta}(t_m) - \mathbf{\hat C}_j \boldsymbol{\beta}(t_m) \|$ 
\end{center}
where each $\mathbf{\hat C}_i$ or $\mathbf{\hat C}_j$ is a $(p \times q)$ coefficient matrix and $\boldsymbol{\beta}(t_m)$ is a vector of $q$ cubic B-spline functions, heuristically, the estimated $\hat {d}_{ij}(t_m)$ converge to $d_{ij}(t_m)$ when $m \to \infty$, that is,
\begin{center}
$\|\hat {d}_{ij}(t_m) - d_{ij}(t_m)\| \to 0$ when $m \to \infty$.
\end{center}
However, the study of the convergence is not the aim of this paper. 

The versatility of FMDS suggests a wide range of potential applications across various fields, including financial market, ecology, psychology, social networks, etc.. In particular, in financial market, FMDS is capable of effectively capturing the dynamic structures present in high-dimensional stock data.
\section{Conclusion}
\noindent In this paper, we proposed Functional Multidimensional Scaling (FMDS), a unified statistical framework that incorporates the advantages of Functional Data Analysis and Classical MDS. By overcoming the inherent limitations of Classical MDS and reducing the computation cost, FMDS offers a powerful alternative for time-varying dissimilarities between objects. In the future work, we plan to conduct comprehensive simulation studies for convergence to identify the feasibility of FMDS and show the analysis of a real-world case to further assess the empirical performance and practical utility of FMDS.
\section*{Acknowledgment}
\noindent The author would like to sincerely thank Dr.Daniel Gervini for the invaluable guidance and support throughout the research.

\bibliographystyle{apalike}

\begin{thebibliography}{}

\bibitem[Ambrosi and Hansohm, 1987]{ambrosi1987dynamischer}
Ambrosi, K. and Hansohm, J. (1987).
\newblock Ein dynamischer ansatz zur repräsentation von objekten.
\newblock In Isermann, H., Merle, G., Rieder, U., Schmidt, R., and Streitferdt, L., editors, {\em Operations Research Proceedings 1986}, pages 425--431, Berlin, Heidelberg. Springer.

\bibitem[Atkinson, 1989]{atkinson1989introduction}
Atkinson, K.~E. (1989).
\newblock {\em An Introduction to Numerical Analysis}.
\newblock John Wiley \& Sons, New York, 2 edition.

\bibitem[Davidon, 1991]{davidon1991variable}
Davidon, W.~C. (1991).
\newblock Variable metric method for minimization.
\newblock {\em SIAM Journal on Optimization}.
\newblock Originally from Argonne National Laboratory, 1959.

\bibitem[de~Boor, 2001]{deboor2001practical}
de~Boor, C. (2001).
\newblock {\em A Practical Guide to Splines}, volume~27 of {\em Applied Mathematical Sciences}.
\newblock Springer, New York, NY, revised edition.
\newblock With 32 figures.

\bibitem[Izenman, 2008]{izenman2008multidimensional}
Izenman, A.~J. (2008).
\newblock Multidimensional scaling and distance geometry.
\newblock In Izenman, A.~J., editor, {\em Modern Multivariate Statistical Techniques: Regression, Classification, and Manifold Learning}, pages 463--504. Springer, New York, NY.

\bibitem[Jäckle et~al., 2016]{Jackle2016TMDS}
Jäckle, D., Fischer, F., Schreck, T., and Keim, D.~A. (2016).
\newblock Temporal mds plots for analysis of multivariate data.
\newblock {\em IEEE Transactions on Visualization and Computer Graphics}, 22(1):141--150.

\bibitem[Kingma and Ba, 2015]{kingma2015adam}
Kingma, D.~P. and Ba, J. (2015).
\newblock Adam: A method for stochastic optimization.
\newblock In {\em Proceedings of the 3rd International Conference on Learning Representations (ICLR)}.

\bibitem[Machado et~al., 2011]{machado2011analysis}
Machado, J. A.~T., Duarte, F.~B., and Duarte, G.~M. (2011).
\newblock Analysis of stock market indices through multidimensional scaling.
\newblock {\em Communications in Nonlinear Science and Numerical Simulation}, 16(12):4610--4618.

\bibitem[Ramsay and Silverman, 2002]{ramsay2002applied}
Ramsay, J.~O. and Silverman, B.~W. (2002).
\newblock {\em Applied Functional Data Analysis: Methods and Case Studies}.
\newblock Springer, New York, NY.

\bibitem[Ramsay and Silverman, 2005]{ramsay2005functional}
Ramsay, J.~O. and Silverman, B.~W. (2005).
\newblock {\em Functional Data Analysis}.
\newblock Springer, New York, NY, 2 edition.

\bibitem[Young and Householder, 1938]{young1938discussion}
Young, G. and Householder, A. (1938).
\newblock Discussion of a set of points in terms of their mutual distances.
\newblock {\em Psychometrika}, 3(1):19--22.

\end{thebibliography}

\end{document}